\documentclass{article}
\usepackage{spconf,amsmath,graphicx}
\usepackage{booktabs}
\usepackage{hyperref}
\copyrightnotice{979-8-3503-0689-7/23/\$31.00~\copyright2023 IEEE}

\title{HiGNN-TTS: Hierarchical Prosody Modeling with Graph Neural Networks for Expressive Long-form TTS}
\name{Dake Guo$^{1}$, Xinfa Zhu$^{1}$, Liumeng Xue$^{2}$, Tao Li$^{1}$, Yuanjun Lv$^{1}$, Yuepeng Jiang$^{1}$, Lei Xie$^{1*}$\thanks{* Corresponding author.}}

\address{$^1$Audio, Speech and Language Processing Group (ASLP@NPU), School of Computer Science, \\ Northwestern Polytechnical University, Xi'an, China\\
$^2$ School of Data Science, The Chinese University of Hong Kong, Shenzhen (CUHK-Shenzhen), China}
\begin{document}

\maketitle
\begin{abstract}

Recent advances in text-to-speech, particularly those based on Graph Neural Networks (GNNs), have significantly improved the expressiveness of short-form synthetic speech. However, generating human-parity long-form speech with high dynamic prosodic variations is still challenging. To address this problem, we expand the capabilities of GNNs with a hierarchical prosody modeling approach, named HiGNN-TTS. Specifically, we add a virtual global node in the graph to strengthen the interconnection of word nodes and introduce a contextual attention mechanism to broaden the prosody modeling scope of GNNs from intra-sentence to inter-sentence. Additionally, we perform hierarchical supervision from acoustic prosody on each node of the graph to capture the prosodic variations with a high dynamic range. Ablation studies show the effectiveness of HiGNN-TTS in learning hierarchical prosody. Both objective and subjective evaluations demonstrate that HiGNN-TTS significantly improves the naturalness and expressiveness of long-form synthetic speech\footnote{Speech samples: \href{https://anonymous-asru.github.io/HiGNN-TTS/}{https://anonymous-asru.github.io/HiGNN-TTS/}}.

\end{abstract}

\begin{keywords}
Expressive long-form TTS, hierarchical prosody modeling, graph neural network, semantic representation enhancing
\end{keywords}
\section{Introduction}
\label{sec:intro}

Text-to-Speech (TTS), aiming to generate human-like speech from text, has achieved dramatically advanced in naturalness with the proliferation of sequence-to-sequence (seq2seq) based neural approaches
~\cite{DBLP:conf/interspeech/WangSSWWJYXCBLA17,DBLP:conf/nips/RenRTQZZL19,DBLP:conf/icml/KimKS21}.
With the growing demand for anthropomorphic human-computer interaction, there has been increasing interest in generating speech with high expressiveness~\cite{Li2021ControllableET,li22h_interspeech,Li2021ControllableCE,10095776}, which exhibits high dynamic range in prosodic variation, including pitch and duration, etc. Human speech is expressive in nature, and proper expression rendering affects overall speech perception, which is essential for many TTS applications, such as newsreaders and voice assistants, etc.
However, there is still a noticeable gap between synthetic speech and human speech in terms of expressiveness, particularly in long-form speech synthesis scenarios like audiobooks. 
Conventional TTS usually works on the prosody modeling of a single sentence. However, long-form speech usually contains multiple semantically coherent sentences, where the prosody of each individual sentence is also affected by its corresponding context.
Therefore, in long-form speech generation, besides ensuring the coherence of the overall rhythm, it is also necessary to model the local fine-grained and global prosody in each sentence as well as the cross-sentence contextual prosody~\cite{DBLP:journals/corr/abs-1909-03965}.


To address sentence-level global prosody, several works have brought attention to the extraction of a global style embedding from a given reference speech which has successfully captured the global style features of speech~\cite{DBLP:conf/icml/Skerry-RyanBXWS18, DBLP:conf/icml/WangSZRBSXJRS18}. In order to synthesize expressive speech without the need of auxiliary reference speech during inference, some methods attempt to obtain prosodic variations directly from text, which is more practical. By incorporating text-only predictions, the Text-Predicted Global Style Token (TP-GST)~\cite{DBLP:conf/slt/StantonWS18} extends the capabilities of GST~\cite{DBLP:conf/icml/WangSZRBSXJRS18}, enabling the generation of style embeddings or style token weights solely based on textual input. 

However, only sentence-level global prosody modeling lacks local fine-grained prosody information within a sentence, such as pauses and emphasis. By aligning the extracted prosody embedding sequence to the phoneme sequence, some works manage to model fine-grained prosody~\cite{DBLP:conf/interspeech/KlimkovRRD19,tan21_interspeech,DBLP:conf/icassp/LeeK19}. Moreover, the tight interplay of prosody and semantic information within sentences has led to a growing focus on leveraging pre-trained language models (such as BERT~\cite{DBLP:conf/naacl/DevlinCLT19}, XLNET~\cite{DBLP:conf/nips/YangDYCSL19}) to enhance fine-grained prosody representation~\cite{DBLP:conf/interspeech/HayashiWTTTL19, DBLP:conf/interspeech/YangZL19}. Further, to fully utilize the tree-structured syntactic information, some works have leveraged graph-based methods to model word-level prosody. Some studies incorporate BERT embeddings and employ Relational Gated Graph Networks (RGGN) to extract semantic information, and consequently enriches prosodic variations and improves expressiveness in the generated speech~\cite{DBLP:conf/interspeech/ZhouSL0B0M22}. Instead of using BERT embedding, other work utilizes the average pooling of hidden layer representations from the acoustic model as node features to enrich prosodic variation~\cite{DBLP:conf/ijcai/YeZ0022}. By capturing the intricate relations between linguistic units and prosodic cues, these methods facilitate the generation of speech with improved naturalness and expressiveness.

 Considering that the prosody in each individual sentence is also affected by its neighboring sentences, some methods improve the naturalness and prosody consistency of the synthetic speech by harnessing contextual information. The method of using context-aware text embeddings obtained by encoding BERT embeddings of continuous multiple sentences was explored to improve the prosody of individual speech~\cite{DBLP:conf/icassp/XuSZZHZ21,10129796}. Additionally, other methods leverage acoustic contextual information to improve the expressiveness of the sentence. One of works utilizes the preceding acoustic feature to improve the prosody coherence and expressiveness~\cite{DBLP:journals/corr/abs-2012-03763}. Furthermore, a few works use both textual context information and previous acoustic prosodic representations to estimate the style embedding of the single sentence~\cite{10095866}, while others utilize separate text and audio context encoders to encode the respective contexts into global prosody representations~\cite{10096247}.
 
Nevertheless, the aforementioned methods focus on modeling prosody on a single scale. While the expressiveness of human speech can be perceived as a compound of hierarchical factors rather than a mono scale. By employing linguistic-aware, prosody-aware, and sentence position networks, ParaTTS~\cite{DBLP:journals/taslp/XueSZX22} achieves paragraph-level speech synthesis with smooth prosody and coherence. A few approaches incorporate a hierarchical context encoder to capture syllable structural features for both inter-phrase and inter-sentence contexts~\cite{DBLP:conf/icassp/LeiZCWKM22}. Another method, known as multi-scale style modeling, focuses on modeling context at cross-sentence, sentence, and subword levels~\cite{DBLP:conf/interspeech/LeiZCH0KM22}. They both obtain a global prosody representation from different scales, thereby improving the expressiveness of the synthetic speech. 
Although tremendous effort has been made to improve expressiveness in both intra-sentence and inter-sentence scales, listeners still feel obvious fatigue with long-form synthetic speech. In this paper, we propose \textit{HiGNN-TTS}, a graph-based hierarchical prosody modeling approach for long-form expressive speech synthesis. Specifically, we expand GNNs with \textit{hierarchical} prosody modeling capacity covering word, sentence, and cross-sentence levels. First, we add a virtual global node in the syntax tree to improve the connections between word nodes. Second, we design a hierarchical graph encoder to explore the hierarchical structural relationship in the text at the word level, sentence level, and cross-sentence context level, which broadens the scope of graph-based speech synthesis from intra-sentence to inter-sentence. Third, we introduce prosody supervision signals from speech and propagate them to each word node of the graph through the message-passing capability of GNN, utilizing prosodic variations from the speech to guide the prosody modeling from the text. 
Ablation studies show that our proposed approach, HiGNN-TTS, is effective in modeling hierarchical prosody. Objective and subjective experiments demonstrate that HiGNN-TTS outperforms several competitive models, achieving more natural and expressive short-form and long-form synthetic speech.

\section{Methodology}
\label{sec:methodology}
\begin{figure}[h]
    \centering
    \includegraphics[width=0.99\linewidth]{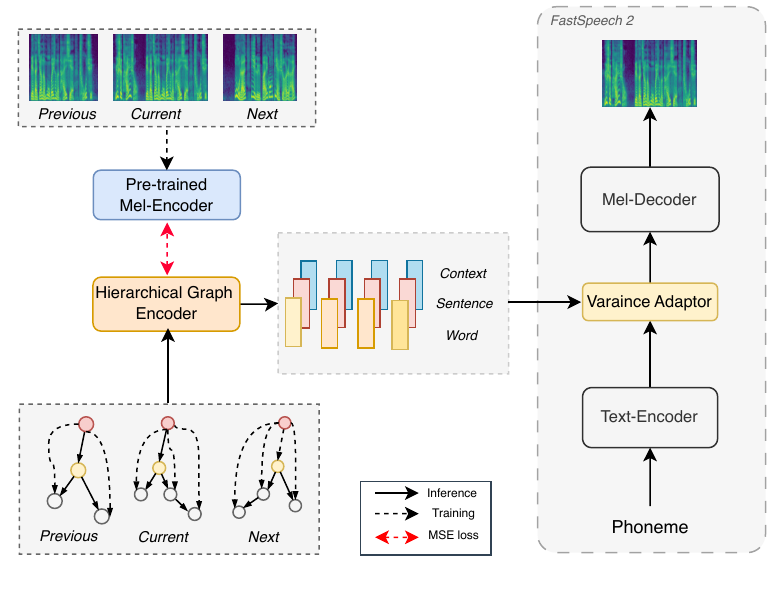}
    \caption{The overall architecture of HiGNN-TTS}
    \label{fig:overview}
\vspace{-10pt}
\end{figure}

As illustrated in Fig.\ref{fig:overview}, the overall architecture of HiGNN-TTS is a Fastspeech2-based model with a newly-introduced hierarchical graph prosody encoder and a pre-trained Mel-encoder.
The hierarchical graph encoder models hierarchical prosody based on syntax graph. It takes in the graphs of previous, current, and next sentences and learns word-level, sentence-level, and cross-sentence context prosody representations. The pre-trained mel-encoder supervises the hierarchical prosody modeling via the acoustic prosody signals, helping capture a wide range of prosodic variations. As the backbone, the acoustic model produces a mel spectrogram from phoneme sequence input on the condition of hierarchical prosody, synthesizing expressive speech.

\subsection{Graph Construction with Syntactic Information }
\label{ssec:Graph}
\begin{figure}[h]
    \centering
    \includegraphics[width=0.75\linewidth]{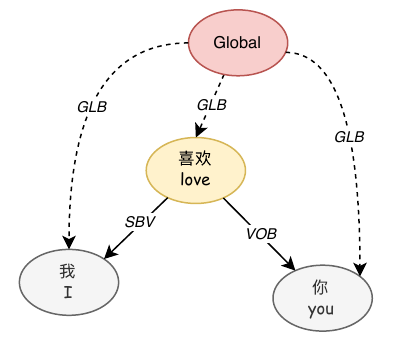}
    \caption{The syntax graph example of  ``I love you." Nodes with different colors represent different constituents: red indicates virtual global nodes, yellow indicates the head of the dependence tree, and gray indicates normal word nodes. GLB, SBV, and VOB represent global relations, subject-verb relations, and verb-object relations, respectively.}
    \label{fig:graph}
    \vspace{-10pt}
\end{figure}

We construct the graph from a syntax tree, which is built by performing dependency parsing at the word level due to the fact that the smallest unit of prosody is a word in Chinese~\cite{DBLP:journals/ijclclp/ChuQ01}~\footnote{In modern Chinese, most words are compounds written with two or more characters.}. Within a syntax tree, bi-lexicalized dependencies are utilized to depict the relationships between words for semantic analysis. Moreover, with syntactic information in the text, we can effectively leverage syntactic relationships to capture prosodic variation. Based on the syntax tree, we introduce a virtual global node to connect all nodes in the tree, forming a syntax graph, as shown in Fig.\ref{fig:graph}. 

The syntax graph can be represented by using a directed graph, defined as $\mathcal{G}=(\mathcal{V}, \mathcal{E})$. Each node is denoted as $v\in \mathcal{V}$, representing a word in the corresponding sentence. Each edge is denoted as $e=(v_i, v_j) \in \mathcal{E}$, representing a specific dependency relation from node $v_i$ to node $v_j$. In this work, we introduce a virtual global node, denoted as $ v_g $, to establish connections to all existing nodes. The connections can be defined as the pair $ e_{v_g, i} = {(v_g, v_i) \mid \forall v_i \in \mathcal{V}} $. The global node improves the model's perception by providing a larger receptive field, enabling it to take into account distant dependencies and effectively capture the prosody patterns present in sentences. 

We initialize graph node features using the BERT embedding extracted from a pre-trained Chinese BERT~\footnote{ \href{https://huggingface.co/bert-base-chinese}{https://huggingface.co/bert-base-chinese}}. For each word node, we first extract BERT embeddings of all Chinese characters in the word and then perform average pooling to get the word nodes' BERT embedding. Additionally, the global node is initialized using BERT CLS embedding.

\subsection{Hierarchical Prosody Modeling based on Graph}
\label{ssec:HPM}
\begin{figure}[h]
  \centering
  \centerline{\includegraphics[width=0.7\linewidth]{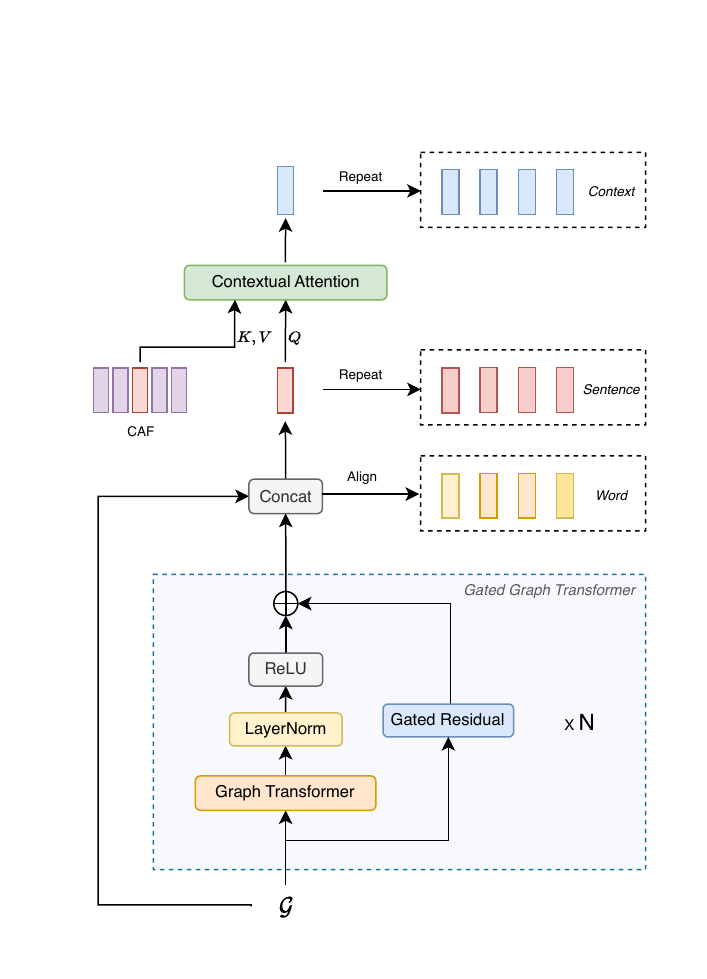}}
    \caption{The architecture of hierarchical graph encoder}
    \label{fig:hge}
    \vspace{-10pt}
\end{figure}

We propose to model hierarchical prosody based on the graph via the hierarchical graph encoder, as shown in Fig.\ref{fig:hge}. The hierarchical graph encoder takes the graphs $\mathcal{G}$ of the previous, current, and next sentences as input to capture the word-level prosody within the current sentence, the sentence-level prosody in the current sentence, and the cross-sentence context from the neighboring sentences.


Firstly, we adopt Gated Graph Transformer~\cite{DBLP:conf/ijcai/ShiHFZWS21} to encode $\mathcal{G}$ into $\mathcal{G^{'}}$, in which each node features are updated with the consideration of correlated node features, enhancing the connections between each word and its surrounding words within a sentence. Specifically, the Gated Graph Transformer contains N layers of Graph Transformer with a gated residual connection, as shown in the bottom part of Fig.~\ref{fig:hge}. Graph Transformer is based on the Transformer model and performs weighted aggregation of node features using a self-attention mechanism. The update of the $i$-th node feature can be formulated by the following equation:
\begin{align}
\mathbf{x}^{\prime}_i & = \mathbf{W}_1 \mathbf{x}_i +
\sum_{j \in \mathcal{N}(i)} \alpha_{i,j} \left(
\mathbf{W}_2 \mathbf{x}_{j} + \mathbf{W}_3 \mathbf{e}_{ij}
\right),
\end{align}
where $\mathbf{x}^{\prime}_i$ represents the updated node feature and $\mathbf{x}_i$ denotes the original node feature. $\mathcal{N}(i)$ denotes the set of neighboring nodes of node $i$, $\mathbf{e}_{ij}$ corresponds to the edge between node $i$ and its neighboring node $j$, and $\alpha_{i,j}$ represents the learnable attention weight between node $i$ and its neighboring node $j$. $\mathbf{W}_1$, $\mathbf{W}_2$, and $\mathbf{W}_3$ are learnable weight matrices applied for linear transformations.

After encoding $\mathcal{G}$ to $\mathcal{G^{'}}$, we fusion $\mathcal{G}$ and $\mathcal{G^{'}}$ via a concatenate operation as the final graph and consumed by the following operations. We believe that the encoded $\mathcal{G^{'}}$ involves acoustic prosody because the hierarchical graph encoder is jointly trained with the acoustic model and constrained by the pre-trained Mel-encoder outputs (as described in Section~\ref{ssec:UPRA}). Hence, the node features in the final fusion graph incorporate not only semantic information in the text but also acoustic prosody in the speech. Similar to GST, we apply a tanh activation to each node in the fusion graph to further enhance the prosody diversity of each node feature. 

Additionally, the introduction of the global node in $\mathcal{G}$ allows the Gated Graph Transformer to connect all nodes in the graph, thus better understanding the overall sentence structure and providing sentence-level prosody. Consequently, in this work, we regard the global node feature in the fusion graph of the current sentence as the corresponding sentence-level prosody representation and the other node features as word-level prosody representation.

To capture cross-sentence contextual prosody, we first concatenate the sentence-level representations of previous, current, and next sentences, forming the aggregated features at the long-form level, referred to as the \textit{Context Aggregation Features} (CAF). And then, we utilize a contextual attention mechanism to learn the cross-sentence prosody, where CAF is used as the key and value, and the current sentence-level representation acts as the query. In fact, the contextual attention mechanism can be considered as a certain type of GNN, in which we treat each sentence representation as a node and the entire concatenated representation as a fully connected graph. By using the attention mechanism, we calculate the attention weights between each sentence and other sentences, effectively aggregating contextual information. This process is similar to massage propagation in GNN.

To facilitate the incorporation of the hierarchical prosody representations learned from the hierarchical graph encoder into the acoustic model, we align each word-level representation to the length of the phoneme sequence of each corresponding word, and repeat the sentence-level and cross-sentence prosody representations, forming the same length as the phoneme sequence of the current sentence.

\subsection{Hierarchical Supervision from Acoustic Prosody}
\label{ssec:UPRA}
In order to capture the prosodic variations in speech during hierarchical prosody modeling, we employ a pre-trained mel-encoder model to supervise the prosody representations learned in the hierarchical graph encoder. The mel-encoder is pre-trained on a multi-speaker multi-style dataset to extract sentence-level acoustic prosody embedding from the mel spectrogram of a given reference speech. To obtain a pure prosody embedding, we decouple the speaker-related information in the speech from the extracted prosody representation by using the Domain Adversarial Training (DAT)~\cite{DBLP:journals/jmlr/GaninUAGLLML16} strategy. Specifically, a speaker classifier with a gradient reversal layer is adopted on the extracted prosody embedding, making the prosody embedding to be speaker indistinguishable.

To supervise hierarchical prosody, speaker-independent acoustic prosody embeddings are extracted from the speech of previous, current, and next sentences, respectively, and then used to directly constrain the corresponding sentence-level prosody representations (i.e., the global node feature) learned in the graph encoder via the mean squared error (MSE) loss. Although the supervision is directly conducted on the sentence-level prosody representation, the cross-sentence prosody representation can also be influenced because it is computed based on the sentence-level prosody representation. Furthermore, with the help of the message-passing capability of GNN, the supervision signals propagate from the sentence-level to word-level nodes, enabling the supervision of acoustic prosody in speech to hierarchical levels. Through hierarchical acoustic prosody supervision, we are able to capture the prosodic variations with high dynamic range comprehensively, thus improving the expressiveness of speech synthesis.


\section{EXPERIMENTS}
\label{sec:exp}

\subsection{Training Setup and Model Details}
\label{ssec:train}
In the experiments, we use an internal multi-speaker audiobook corpus. The corpus contains roughly 8 hours of recordings collected from two professional Mandarin native speakers. The dataset has a total of 7,016 audio clips, 90\% of which of the clips are used for training, and the rest of the clips are used for validation and testing. For feature extraction, 80-dimensional mel-spectrograms are extracted with a 24kHz sampling rate. The frame size is set to 1,024, and the hop size is set to 256. The auxiliary pitch contour is extracted through WORLD vocoder~\cite{209840090ed54721bc71de7d80ee9658}. The phoneme duration is obtained through an HMM-based force alignment model. We use DDParser~\cite{zhang2020practical} as our dependency parser~\footnote{\href{https://github.com/baidu/DDParser}{https://github.com/baidu/DDParser}}. We train all models up to 400k steps on two 3090 GPUs with a batch size of 8. In addition, we use a DSPGAN~\cite{10095105}which is trained on an internal studio-quality dataset with 251 hours from 308 speakers, to convert the generated mel-spectrogram into waveforms.

 The Fastspeech2 related model configuration is consistent with the original paper~\cite{DBLP:conf/iclr/0006H0QZZL21}. The Gated Graph Transformer consists of 4 layers of Graph Transformer with a gated residual connection. Notably, the dimensions of node features and edge embeddings are both set to 256. Within the graph transformer, the attention mechanism employs 4 heads with a dropout rate of 0.1. 

\subsection{Compared Methods}
\label{ssec:train}

To evaluate the performance of HiGNN-TTS, three baseline models are implemented for comparison:
\begin{itemize}
    \item \textbf{FS2-BERT:} a FastSpeech2-based model, which considers context information by using cross-sentence BERT embedding~\cite{DBLP:conf/interspeech/HayashiWTTTL19}.
    \vspace{-5pt}
    \item \textbf{HCE:} a FastSpeech2-based model, which utilizes Hierarchical Context Encoder (HCE) to predict the sentence-level style embedding from the hierarchical context information ~\cite{DBLP:conf/icassp/LeiZCWKM22}.
    \vspace{-5pt}
    \item \textbf{ATCE:} a FastSpeech2-based model with Acoustic and Text Context Encoders (ATCE), which uses both text and audio context to obtain context prosody representations~\cite{10096247}.
    \vspace{-5pt}
\end{itemize}

\subsection{Objective Evaluation}
\label{ssec:oe}
We evaluate the model's performance on the reserved short-sentence testing set with the number of 50 using several widely used objective evaluation metrics as in prior studies\cite{DBLP:conf/iclr/0006H0QZZL21}. These metrics include the root mean square error (RMSE) for pitch, mean square error (MSE) for the duration, and mel cepstral distortion (MCD). To calculate the RMSE for F0 and MCD, we employ dynamic time warping (DTW)~\cite{DTW} to align the predicted mel-spectrogram with the ground-truth mel-spectrogram.

\begin{figure}[h]
    
  \centering
  \hspace{-10pt}
  \centerline{\includegraphics[width=1.0\linewidth]{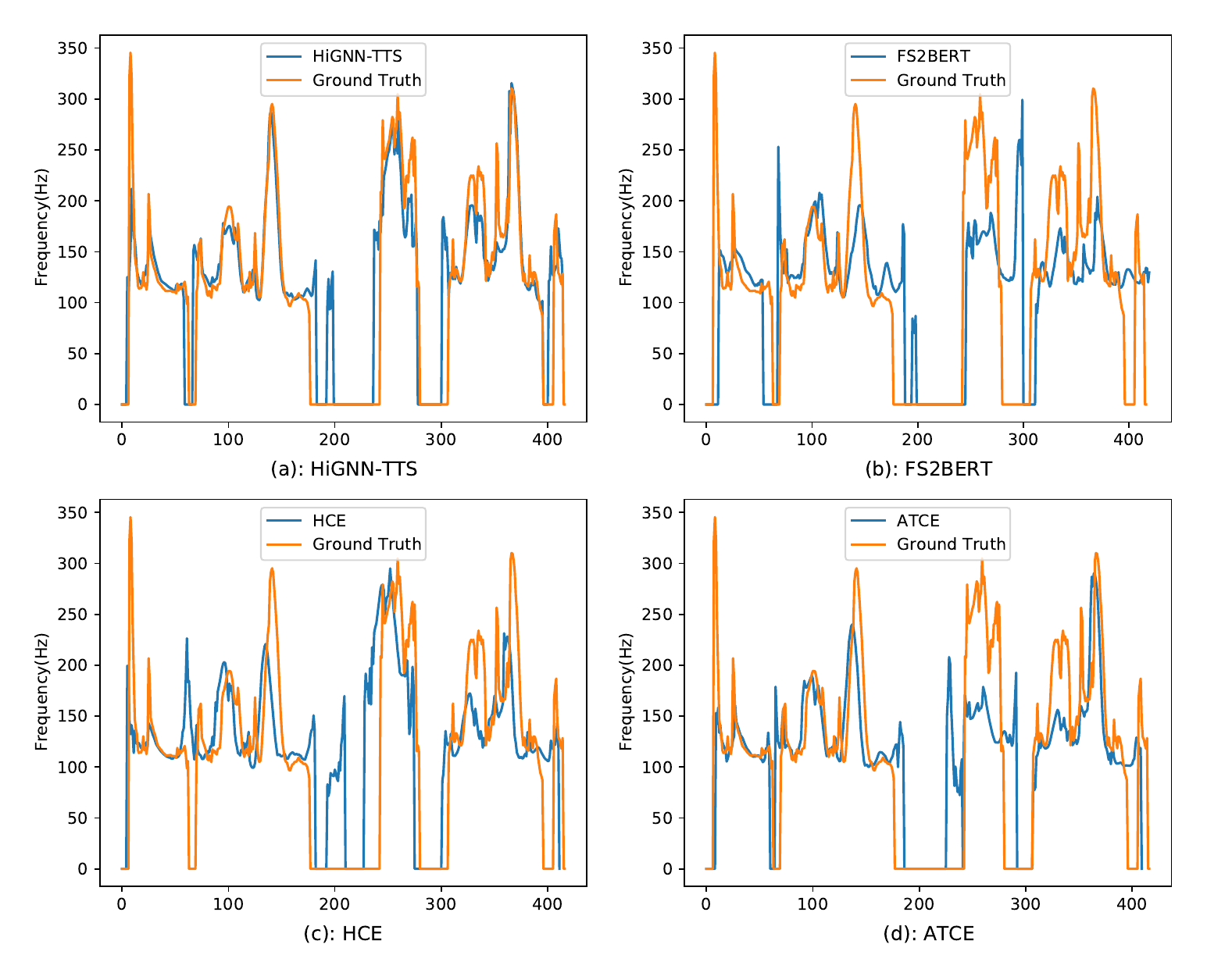}}
    \caption{F0 contours of the same speech synthesized by different models.}
    \label{fig:pitch}
\end{figure}

The objective results are shown in Table \ref{obj}. HiGNN-TTS achieves superior performance over the three baselines on all objective evaluation metrics. Lower MCD and Duration MSE indicate that HiGNN-TTS obtains better mel and duration prediction and achieves more natural speech synthesis. Notably, we observe a significant degradation in the F0 RMSE of HiGNN-TTS compared to other metrics. This can be attributed to that our hierarchical modeling approach with the supervision of acoustic prosody effectively captures local prosodic variations with high dynamics. Furthermore, we present F0 contours of the same speech synthesized by different models in Figure \ref{fig:pitch}. It reveals that the F0 pattern of our model is close to that of the ground truth recording, indicating that our approach is effective in capturing hierarchical prosodic variations.
\begin{table}[t]
    \centering
    \caption{Objective evaluations for different models.}
    \label{obj}
    \vspace{5pt}
\begin{tabular}{lccc}
\toprule
\multicolumn{1}{c}{\textbf{Model}} & \textbf{MCD $\downarrow$}   & \textbf{Duration MSE $\downarrow$} & \textbf{F0 RMSE $\downarrow$} \\  \midrule
FS2-BERT                           & 4.283          & 0.1421                & 51.1192          \\
HCE                                & 4.073          & 0.1197                & 41.2083          \\ 
ATCE                               & 4.123          & 0.1243                & 42.4219          \\
HiGNN-TTS                            & \textbf{3.979} & \textbf{0.1083}       & \textbf{36.6086} \\ 
\bottomrule
\end{tabular}
\end{table}
\vspace{5pt}
\vspace{-10pt}
\subsection{Subjective Evaluation}
\label{ssec:se}
We conducted Mean Opinion Score (MOS) tests to evaluate the expressiveness and naturalness of the synthetic speech in a similar way to \cite{DBLP:conf/interspeech/LeiZCH0KM22,DBLP:conf/icassp/LeiZCWKM22}. Moreover, we evaluate 20 short-form sentences (S-MOS) and 10 long-form sentences (L-MOS). S-MOS consists of individual sentences with context, each lasting about 5 seconds. And L-MOS is constructed by combining multiple sentences with a duration of approximately 40 seconds. In each MOS test, a group of 20 native Chinese Mandarin listeners is asked to listen to synthetic speeches and rate them on a scale from 1 to 5 with a 0.5-point interval. 
\vspace{-0pt}
\begin{table}[h]
    \centering
    \caption{The S-MOS and L-MOS of different models with 95\% confidence intervals.}
    \label{mos}
    \vspace{5pt}
\begin{tabular}{ccc}
\toprule
\textbf{Model}    & \textbf{S-MOS $\uparrow$} & \textbf{L-MOS $\uparrow$} \\  \midrule
FS2-BERT &    3.52 $\pm$ 0.124&   3.34 $\pm$ 0.113 \\ 
HCE      &   3.85 $\pm$ 0.116  &   3.46 $\pm$ 0.101\\ 
ATCE     &    3.91 $\pm$ 0.102 &    3.44 $\pm$ 0.121\\ 
HiGNN-TTS  &   \textbf{4.09 $\pm$ 0.098}  &  \textbf{3.86 $\pm$ 0.113}  \\ \bottomrule
\end{tabular}
\vspace{-3pt}
\end{table}

As presented in Table \ref{mos}, our proposed method, HiGNN-TTS, achieves the best S-MOS of 4.09 and L-MOS of 3.86. By effectively leveraging semantic information, HiGNN-TTS achieves superior performance compared to the other models. Moreover, it is worth noting that, due to potential listener fatigue when listening to long-form speech, the L-MOS score is consistently lower than the S-MOS score. Even so, our hierarchical prosody modeling approach still exhibits remarkable superiority over all baseline models in long-form speech. The higher S-MOS and L-MOS scores demonstrate our model's capability to model hierarchical prosody and produce more expressive speech.

\subsection{Ablation Study}
\label{sec:as}

We perform ablation studies by removing acoustic prosody supervision, contextual attention, and global node (sentence-level representation) in HiGNN-TTS. Then we conduct comparison mean opinion score (CMOS) tests of expressiveness and naturalness on both long-form and short-form speech. 

\vspace{-15pt}
\begin{table}[h]
    \centering
    \caption{Results of ablation studies.}
    \label{as}
    \vspace{5pt}
\begin{tabular}{lc}
\toprule
\textbf{Model}    & \textbf{CMOS}  \\ \midrule
HiGNN-TTS &         0 \\    
~- speech supervision     &      -0.931   \\ 
~- contextual attention     &     -0.510    \\ 
~~~- global node  &        -0.483 \\ \bottomrule
\end{tabular}
\end{table}

The results, listed in Table \ref{as}, show that all the designs in HiGNN-TTS yield a significant positive effect on the expressiveness of generated speech. Specifically, When removing the acoustic prosody supervision strategy from HiGNN-TTS, we find it results in the largest CMOS decrease. This indicates that acoustic prosody supervision helps hierarchical graph prosody encoder to capture more large dynamic range of prosodic variations. Moreover, removing contextual attention also leads decrease in CMOS because of the lack of cross-sentence context information. Based on the removal of contextual attention, we further remove the global node in the syntax graph, and it results in CMOS reductions due to the absence of the overall rhythm in the individual sentence.

\section{CONCLUSION}
\label{sec:conclusion}

In this work, we propose HiGNN-TTS, a graph-based hierarchical prosody modeling for expressive long-form speech synthesis. We modify the syntax tree by the addition of a virtual global node, which enhances the relationships with the word nodes.
Additionally, a hierarchical graph encoder is designed to capture the hierarchical structural relationship in the text at the word level, sentence level, and cross-sentence context level, which broadens the scope of graph-based speech synthesis from intra-sentence to inter-sentence. Furthermore, we introduce the signal of acoustic prosody in the speech to hierarchically supervise the prosody learned from the text in the graph. Ablation studies show the effectiveness of the hierarchical graph encoder and hierarchical supervision from acoustic prosody. Experimental results demonstrate that our proposed HiGNN-TTS significantly improves prosodic variations and generates more natural and expressive short-form and long-form speech.

\bibliographystyle{IEEEbib}
\bibliography{refs}

\end{document}